# A Computationally Efficient, Robust Methodology for Evaluating Chemical Timescales with Detailed Chemical Kinetics.

S. M. Aithal

Computational Sciences Division, Argonne National Laboratory,
9700 S. Cass Ave., Argonne, IL 60439, USA

Phone : 630-252-3164, e-mail: aithal@cels.anl.gov

**Abstract:** Turbulent reacting flows occur in a variety of engineering applications such as chemical reactors and power generating equipment (gas turbines and internal combustion engines). Turbulent reacting flows are characterized by two main timescales, namely, flow timescales and chemical (or reaction) timescales. Understanding the relative timescales of flow and reaction kinetics plays an important role, not only in the choice of models required for the accurate simulation of these devices but also their design/optimization studies. There are several definitions of chemical timescales, which can largely be classified as algebraic or eigenvalue-based methods. The computational complexity (and hence cost) depends on the method of evaluation of the chemical timescales and size of the chemical reaction mechanism. The computational cost and robustness of the methodology of evaluating the reaction times scales is an important consideration in large-scale multi-dimensional simulations using detailed chemical mechanisms. In this work, we present a computational efficient and robust methodology to evaluate chemical timescales based on the algebraic method. Comparison of this novel methodology with other traditional methods is presented for a range of fuel-air mixtures, pressures and temperatures conditions. Additionally, chemical timescales are also presented for fuel-air mixtures at conditions of relevance to power generating equipment. The proposed method showed the same temporal characteristics as the eigenvalue-based methods with no additional computational cost for all the cases studied. The proposed method thus has the potential for use with multidimensional turbulent reacting flow simulations which require the computation of the Damkohler number.

**Nomenclature**

| | |
|---|---|
| $E_i$ | Activation energy in the rate constant of the 'i$^{th}$' reaction (J/mole) |
| *i* | reaction index |
| *I* | Total number of reactions |
| *k* | species index |
| *K* | Total number of species |
| $k_{fi}$ | forward rate constant of the $i^{th}$ reaction (units depend on reaction) |
| $k_{ri}$ | reverse rate constant of the $i^{th}$ reaction (units depend on reaction) |
| $q_i$ | rate of progress of the $i^{th}$ reaction (moles/cm$^3$ sec) |
| $R_u$ | universal gas constant (J/mole-K) |
| t | time (sec) |
| T | temperature (K) |
| $\overline{W}$ | Average molecular weight of the mixture (g/mole) |
| $W_k$ | Molecular weight of the $k^{th}$ species (g/mole) |
| $[X_k]$ | Molar concentration of the $k^{th}$ species (moles/cm$^3$) |
| $Y_k$ | mass fraction of the $k^{th}$ species |

**Greek Symbols**

| | |
|---|---|
| $\rho$ | density (g/cm$^3$) |
| $\tau$ | characteristic chemical time scale (sec) |
| $v_{ki}$ | Stoichiometric coefficient of the $k^{th}$ species in the $i^{th}$ reaction |
| $v'_{ki}$ | Stoichiometric coefficient of the $k^{th}$ reactant species in the $i^{th}$ reaction |
| $v''_{ki}$ | Stoichiometric coefficient of the $k^{th}$ product species in the $i^{th}$ reaction |
| $\dot{\omega}_k$ | net production rate of species $k$ (moles/cm$^3$sec) |

**Abbreviations**

| | |
|---|---|
| IRRTS | Inverse Reaction Rate Time scale |
| RTS | Ren Time scale |
| RPTS | Ren Product Time scale |
| IETS | Inverse EigenValue Time Scale |

# 1 Introduction

Turbulent reacting flows occur in a wide variety of engineering applications such as gas turbines and internal combustion (IC) engines. Understanding the interaction between the flow turbulence and chemical kinetics is important in the design, optimization and reliable operation of these devices. For instance, flow and chemistry interactions are useful in addressing practical problems such as flame blow-off while making aircraft engines smaller and lighter. These interactions are also important in improving combustion efficiency, use of leaner fuel-air mixtures and using a more diverse set of fuels for conventional power generation devices. Understanding turbulence-chemistry interaction is thus key, not only in understanding the physics of combustion but also in making judicious engineering choices for optimal equipment design.

Turbulence-chemistry interaction depends on two main timescales, namely, the flow timescales and the chemistry timescales. The Damköhler (Da), defined as the ratio of mixing/flow timescales to chemical timescales is an important parameter that characterizes the behavior of the reacting flow system based on flow/turbulence and chemical kinetics ($\tau_f/\tau_c$) [1]. While the definitions of the flow timescales are well-defined, namely integral timescales and Kolmogorov timescales, there are several different definitions of chemical timescales with varying degrees of complexity. The chemical timescale can be computed using two methods (i) algebraic methods (ii) eigenvalue-based methods. Algebraic methods define chemical timescales based on reaction-rate constants, the net-production rate of a species and species mass-fractions [2-4]. Eigenvalue-based methods define the chemical timescales based on the Jacobian describing the reacting flow system [5-10]. Reference [11] has a detailed discussion on both the algebraic and eigenvalue-based methods.

The main objective of this paper is to present a novel, robust and computationally efficient algebraic method to compute chemical timescales for complex chemical mechanisms. In addition to providing insight into the chemical kinetic timescales, this approach can be used in multi-dimensional reacting flow simulations where the turbulence-chemistry interactions are modeled using the Eddy Dissipation Concept (EDC) model as in [11].

This paper is organized as follows. Section 2 discusses the governing equations describing a constant pressure, adiabatic combustion system and reviews common methods used to compute chemical timescales for such systems. Section 3 discusses the proposed method and discusses its advantages compared to currently used methods. Section 4 presents validation of the method and the importance of tight numerical tolerances in computing the species mass fractions. Section 4 further presents application of this method to various case studies under different thermodynamics constraints (isothermal and adiabatic), fuel-air mixtures, and initial conditions of temperature and pressures. Section 5 briefly summarizes the main findings of this work.

## 2 Governing Equations and chemical timescales definitions

### 2.1 Governing equations

A constant pressure, adiabatic combustion system can be described by the coupled solution of the mass and energy conservation equations as shown in Eq (1) and (2), respectively.

$$\frac{dY_k}{dt} = \frac{\dot{\omega}_k}{\rho} W_k \tag{1}$$

$$\frac{dT}{dt} = -\frac{1}{\rho \bar{c}_p} \sum_{k=1}^{K} h_k \dot{\omega}_k W_k \tag{2}$$

where,

$$\dot{\omega}_k = \sum_{i=1}^{I} v_{ki} q_i \tag{3}$$

and

$$q_i = k_{f_i} \prod_{k=1}^{K} [X_k]^{v'_{ki}} - k_{r_i} \prod_{k=1}^{K} [X_k]^{v''_{ki}} \tag{4}$$

$$v_{ki} = v''_{ki} - v'_{ki} \tag{5}$$

$$K_{fi} = A_i T^{\beta_i} exp\left(\frac{-E_i}{R_u T}\right) \tag{6}$$

where '$i$' is the reaction index.

If the reacting system $f$ has $K$ species, the time evolution of the mass fractions can be expressed in matrix form as

$$f = \begin{bmatrix} Y_1 \\ Y_2 \\ . \\ . \\ . \\ Y_K \end{bmatrix} = \begin{bmatrix} f_o(\rho), Y_1 Y_2, \dots Y_K \\ f_1(\rho), Y_1 Y_2, \dots Y_K \\ . \\ . \\ . \\ f_N(\rho), Y_1 Y_2, \dots Y_K \end{bmatrix} \tag{7}$$

and the Jacobian $J$, representing the system can be written as

$$J = \begin{bmatrix} \frac{\partial Y_1}{\partial y} \\ \frac{\partial Y_2}{\partial y} \\ \cdot \\ \cdot \\ \cdot \\ \frac{\partial Y_K}{\partial y} \end{bmatrix} = \begin{bmatrix} \frac{\partial Y_1}{\partial Y_1} & \frac{\partial Y_1}{\partial Y_2} & \cdots & \frac{\partial Y_1}{\partial Y_K} \\ \frac{\partial Y_2}{\partial Y_1} & \frac{\partial Y_2}{\partial Y_2} & \cdots & \frac{\partial Y_2}{\partial Y_K} \\ \cdot & \cdot & & \cdot \\ \cdot & \cdot & & \cdot \\ \cdot & \cdot & & \cdot \\ \frac{\partial Y_K}{\partial Y_1} & \frac{\partial Y_K}{\partial Y_2} & \cdot & \frac{\partial Y_K}{\partial Y_K} \end{bmatrix} \qquad (8)$$

## 2.2 Chemical timescale definitions

As stated above, algebraic methods define timescales as functions of the net production rate (the RHS of Eq. (1)) along with the species mass fractions or reaction rate constants. Some of the common definitions of timescales based on the algebraic method are shown below and discussed in this work.

Inverse Reaction Rate Time Scale (IRRTS) is defined as

$$\tau_{IRRTS} = \min_{i \in I} \frac{1}{\left|\left(\frac{\overline{W}}{\rho}\right) q_i\right|} \qquad (9)$$

where $I$ is the maximum number of reactions in the mechanism. The rate of progress of reaction has units of moles/(volume-time) and hence must be multiplied by $\overline{W}/\rho$ to yield dimension of 1/time. Thus, the time-constant of a system is defined based on the fastest reaction rate. This definition places equal importance on all reactions in a system. Since the reaction rate is a product of the reaction rate constants ($k_f/k_r$) and the species concentrations as shown in Eq (4), the temporal variation of the timescales of a system can vary by several orders of magnitude from the initial to the final state. In combustion systems, the concentration of certain species such as the fuel and/or oxidizer change from high initial values to near zero at the final stages leading to large temporal variations in the timescales of the system.

The Ren Time Scale (RTS) is defined as

$$\tau_{RTS} = \min_{k \in K} \left( \frac{Y_k}{\left(\frac{W_k}{\rho}\right) \dot{\omega}_k} \right) \qquad (10)$$

For all species with $\dot{\omega}_k < 0$. Since $Y_k > 0$ and $\dot{\omega}_k < 0$, the computed values of $\tau_{RTS}$ from Eq. 10 is always negative. The absolute value of the computed maximum value of $\tau_{RTS}$ (least negative value) is reported as the chemical timescale for the RTS method.

Similar to the RTS timescale, the Ren Product Time Scale (RTPS) is defined as

$$\tau_{RPTS} = \min_{k \in K} \left( \frac{Y_k}{\left(\frac{W_k}{\rho}\right)\dot{\omega}_k} \right) \qquad (11)$$

For $\dot{\omega}_k > 0$.

As shown in Eq (3), the net rate of production of species $k$, $\dot{\omega}_k$, is based on the production/depletion of the species due to all reactions in a mechanism. The net rate of production of a species depends on the reaction rate constants and molar concentrations of various species which changes continuously with time as the system proceeds from the initial state (temperature & composition) to the final state. The temporal variation of temperature computed using Eq (2) is used to compute the reaction rate constants as shown in Eq (6). One of the main drawbacks of using timescales based on algebraic methods is that they rely on mass fractions and $\dot{\omega}_k$. As the system approaches steady state, the RHS of Eq (1) approaches zero. Similarly, the steady-state values of many intermediate species in the mechanisms would also approach zero. Timescales based on algebraic methods can yield physically meaningless values when $\dot{\omega}_k$ and/or $Y_k$ are in the numerical noise (values such as $10^{-40}$, $10^{-75}$ etc.). The computed values of $\dot{\omega}_k$ and $Y_k$ are also strongly affected by the numerical tolerances and hence can yield misleading results.

There are several definitions of timescales based on the Jacobian and/or eigen value(s) describing the reacting flow systems. Inverse Jacobian Time Scale (IJTS), System Progress Time Scale (SPTS), Inverse EigenValue Time Scale (IETS) and EigenValue Time Scale (EVTS) are some examples. Reference [11] describes these methods in detail. The large computational cost of computing the Jacobian matrix with sufficient numerical accuracy is the main drawback of eigen value-based methods for computing timescales. While it may seem that the Jacobian computed for implicit time-marching might be used to compute the chemical timescales, this is not the case. Most time-marching schemes compute the Jacobian matrix numerically using methods such as finite-differencing. The accuracy of the Jacobian matrix for time-marching is not of high importance. The use of an approximate Jacobian for time-marching may affect the convergence rate but not the accuracy of the solution. Many numerical schemes exploit this consideration to compute/update the Jacobian matrix once every few time-steps to reduce the overall time to solution. Using an approximate Jacobian will however not yield accurate chemical timescales in eigenvalue-based methods. The Jacobian matrix can be computed using analytic expressions to avoid loss of accuracy due to numerical differentiation, but this process can be tedious and time-consuming. The computational cost will also be prohibitive for multi-dimensional flow simulations where it is necessary to compute the timescale for each grid-point/cell at each time-step. This problem is further exacerbated for multidimensional simulations using detailed chemical kinetics with tens to hundreds of species. Hence, eigenvalue-based timescale computations are not practical for multidimensional CFD simulations. In this work, we will discuss the IETS method as a representative eigenvalue-based method for comparison with the algebraic methods.

The IETS method defines the timescale as shown below

$$\tau_{IETS} = \min_{k \in K}\left(\frac{1}{|\lambda_k|}\right) \tag{12}$$

## 3 Proposed method

It has been pointed out that defining chemical timescales based on the species concentrations and net production rate can lead non-physical values and/or behavior. Reference [11] discusses a case where a single-step reaction yields two distinct timescales, which is non-physical since a single-step reaction is characterized by a single timescale. Drawbacks associated with the timescales computed using various algebraic methods can be remedied if the RHS of Eq (1) can be written as

$$\frac{\omega_k}{\rho} W_k = P_k Y_k + C_k = \frac{Y_k}{\tau_k} + C_k; \quad P_k = \frac{1}{\tau_k} \tag{13}$$

In the above equation, $C_k$ represents a collection of all terms in $\omega_k$ that do not include $Y_k$. Based on this description of the RHS, Eq (1) could then be written as

$$\frac{dY_k}{dt} = P_k Y_k + C_k = \frac{Y_k}{\tau_k} + C_k \tag{14}$$

The term $\tau_k$ has time-units (seconds) and the absolute value of $\tau_k$ can be considered as a time constant of the species $k$. Thus, a reaction mechanism with 'K' species will have 'K' different chemical timescales. The extremum values of $|\tau_k|$ represent species with the fastest and slowest kinetics of a system as it proceeds towards steady state. The fastest times scales (lowest value of $|\tau_k|$) is considered to be the chemical timescale. Computation of the RHS of Eq (1) needed to describe the time evolution of the species mass-fraction, also yields the time evolution of the time constants describing each species, at no extra computational cost. The numerical issues associated with the net production rate tending to zero at steady state does not affect the timescale since neither $\tau_k$ nor $C_k$ approach zero even if the net production rate tends to zero. Thus, the proposed method is computationally inexpensive and numerically robust.

## 4 Results and discussions

In this section we present validation of the in-house solver developed to compute the net production rate as discussed in Section 3. The results of the in-house solver are compared to those predicted by Cantera (version 2.5) [12]. We also include the effect of numerical tolerances on the algebraic timescales. The validated solver with the correct numerical tolerances is used to examine the time scale for a series of case studies under perfectly stirred reactor (PSR) conditions.

## 4.1 Code validation

We present validation results for the case of oxidation of CO to $CO_2$ under isothermal conditions reported in Ref. [11]. The initial mixture consists of 2 moles of CO, 1 mole of $O_2$ (stoichiometric mixture) and 0.5 moles of $H_2O$ at a pressure of 1 atm and 1500K. The simulations are time-marched to a final time of 10 milliseconds ($10^{-2}$ sec) using the in-house solver and Cantera (version 2.5.1) [12] with the GRI 3.0 mechanism [13]. Figure 1 shows that the temporal variations of mass fractions of CO and $CO_2$ obtained using Cantera and the in-house solver are in very good agreement. Figure 1 also shows that CO is rapidly oxidized to $CO_2$ within about 100 microseconds (from $10^{-5} < t < 10^{-4}$). For $2 \times 10^{-4} < t < 10^{-2}$ seconds, the change in the $CO_2$ mass fraction is negligible.

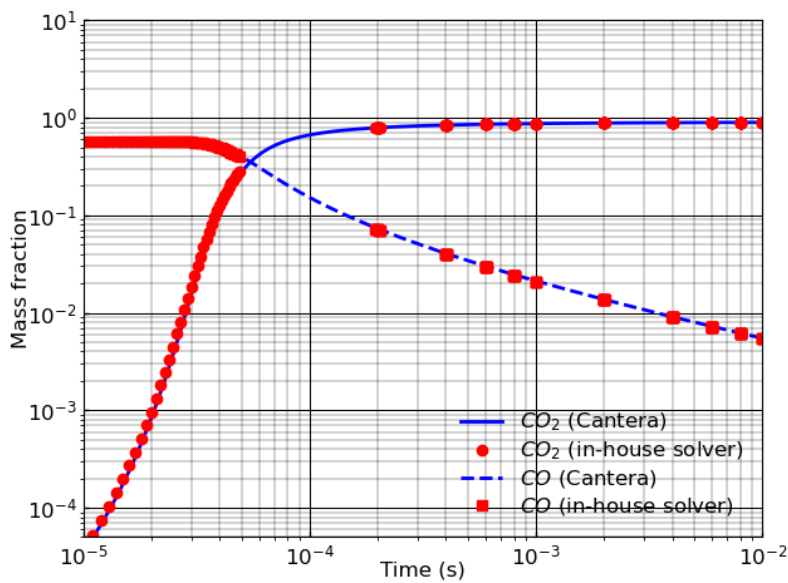

Figure 1: Comparison of species mass fraction between in-house solver and Cantera.

Figure 2 shows very good agreement between the IRRTS chemical time scale obtained using Cantera and the in-house solver with the results reported in Ref. [11].

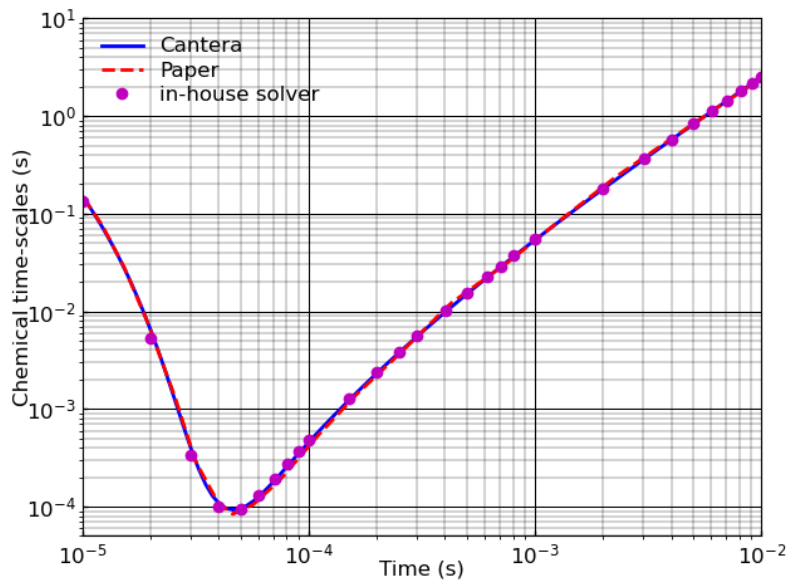

Figure 2: Comparison of the IRRTS chemical timescales (s) predicted by Cantera and in-house solver (isothermal oxidation of CO at 1500K and 1 atm).

## 4.2 Effect of numerical tolerances

Algebraic methods such as RPTS and RTS are defined based on species concentrations and the net production rate ($\dot{\omega}_k$) as discussed in Equations [9-11] shown in Section 2. As the system approaches steady state the net production term tends to zero. The species concentrations and net production rates of trace species and radicals which are formed and destroyed quickly during the combustion process can be very small ($10^{-50} < \dot{\omega}_k < 10^{-20}$), hence very high numerical accuracy of the species concentrations is required during the time-marching to accurately compute the chemical time constants using the RTS and RPTS methods. To ensure that the net production rates and the species mass fractions are computed with sufficient accuracy, very low values of relative and absolute tolerances are required. In this work, we have used the most stringent values of relative and absolute tolerances allowable, namely, an absolute tolerance of $10^{-21}$ and a relative tolerance of $10^{-16}$ in both the in-house solver and Cantera. Less stringent tolerances (such as using an absolute tolerance of $10^{-6}$) do not impact the concentration of major species but can impact the chemical timescale evaluations. Figure 3 and Figure 4 shows the RTS and RPTS time scale computations conducted with the most stringent tolerance values stated above ($10^{-16}$ & $10^{-21}$) along with the same computations conducted with less stringent tolerance criterion (absolute tolerance set to $10^{-6}$). It can be seen that in Figure 3 and Figure 4, the chemical timescales obtained using a an absolute tolerance of $10^{-6}$ is oscillatory in nature.

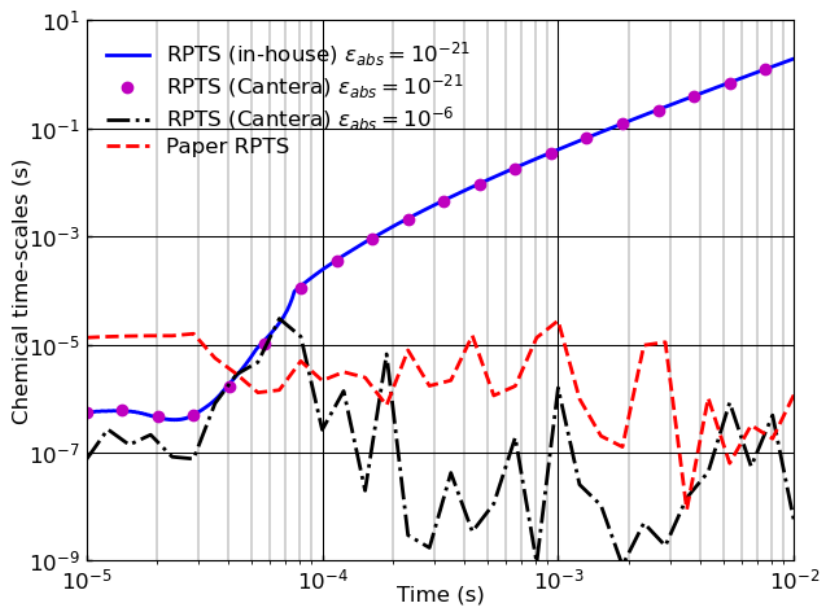

Figure 3: Effect of tolerances on RPTS (isothermal oxidation of CO at 1500K and 1 atm)

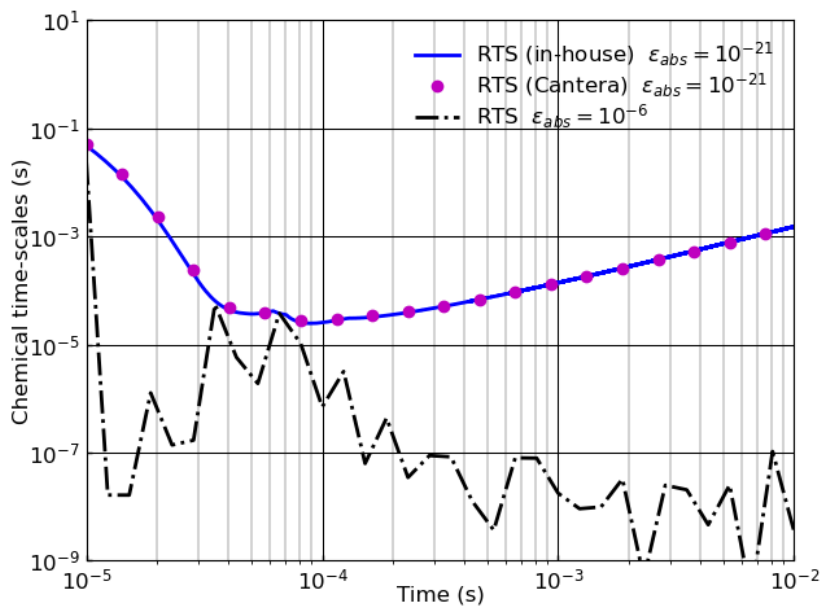

Figure 4: Effect of tolerance on RTS (isothermal oxidation of CO at 1500K and 1 atm)

Reference [11] reports a similar oscillatory chemical timescale for the RTS and RPTS methods (see Figure 4 in [11]). The authors ascribe this oscillatory nature to rapidly changing net production rates of some radical species. The authors of Ref. [11] also do not explicitly state the relative and absolute tolerances used in their Cantera simulations. A careful study of the species determining the chemical timescales for the RTS and RPTS methods with tight and less stringent numerical tolerances showed some interesting results. The main species deciding the chemical timescales using the RTS and RPTS methods with tight numerical tolerances ($10^{-16}$ & $10^{-21}$) are shown in Figure 5.

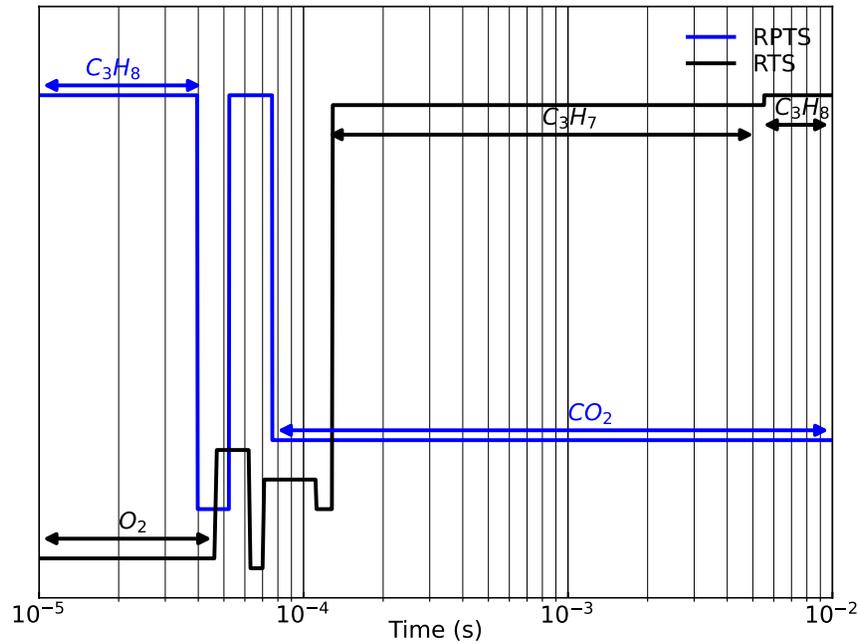

Figure 5: Main species controlling the time scale for RPTS and RTS methods (absolute tolerance = $10^{-21}$ and relative tolerance = $10^{-16}$) for isothermal oxidation of CO at 1500K and 1 atm.

For the RPTS method (blue line), after t >~ $5 \times 10^{-5}$ when CO is rapidly oxidized to $CO_2$, the chemical timescale is determined by $CO_2$ since it is the species with the largest net production rate. The RTS method (black line) determines the chemical timescales based on species that are depleted ($\dot{\omega}_k < 0$) shows that after the oxidation to CO to $CO_2$ is complete, timescale determining species are $C_3H_7$ and $C_3H_8$.

Parametric studies of simulations conducted with relative tolerance $\geq 10^{-15}$ and absolute tolerance $\geq 10^{-18}$, showed that the chemical timescales were oscillatory in nature, for both the RTS and RPTS methods. For these simulations with a less stringent tolerance criteria, the instantaneous chemical timescales were determined by short-lived trace species such as CH, $CH_2CHO$, $CH_3OH$, C, $CH_3O$ etc. whose species concentrations varied between $10^{-20}$ and $10^{-40}$. Since different trace species with low values of $Y_k$ and $\dot{\omega}_k$ determined the timescales at various time-instants the corresponding chemical timescales were highly oscillatory in nature as shown in Figure 3 and Figure 4 (and reported in Ref. [11]). From this study, it is clear that the oscillatory nature of chemical timescales of the RTS and RPTS methods are numerical artifacts due to inadequate convergence tolerances. Hence, it is very important to have very stringent numerical tolerances to obtain accurate (non-oscillatory) chemical timescales using the RTS and RPTS methods.

### 4.3  Case studies:

In this section we present comparison of the proposed new method of computing chemical timescales with other traditional methods, namely, RTS, RPTS and IRRTS (algebraic) and IETS (eigenvalue-based) for several cases. In addition to the simple system of isothermal oxidation of CO discussed in Ref. [11], we present the combustion of practical fuels such as hydrogen and methane. Since most engineering applications do not involve combustion under isothermal conditions, we present the combustion of these fuels under adiabatic conditions as well. The following cases will be discussed for both isothermal and adiabatic conditions.

1. Oxidation of stoichiometric $CO/O_2$ mixture (1500K, 1 atm)
2. Oxidation of stoichiometric $H_2/O_2$ mixture (1500K, 1 atm)
3. Oxidation of stoichiometric $CH_4/O_2$ mixture (1500K, 1 atm)
4. Oxidation of lean $CH_4/O_2$ mixture (1500K, 1 atm, equivalence ratio ($\varphi$) = 0.8)

In addition to these cases, we will also discuss cases for conditions at the "start-of-ignition" (SOI) in engines fueled with natural gas. In a typical reciprocating engine using natural gas as a fuel, the cylinder pressure at SOI is about 25 atm and the fuel-air mixture is at about 750K at the time of sparking. For these conditions, we will study a stoichiometric $CH_4/O_2$ mixture and a lean $CH_4/O_2$ mixture with $\varphi = 0.8$.

*4.3.1    Isothermal cases:*

The time-evolution of a chemical system under constant temperature (isothermal) conditions is accomplished by the coupled solution of the system of equations for the species evolution described by Eq (1) and setting the RHS of Eq (2) to zero (implying no change in temperature with time). Figure 6 shows the four case studies for $0 < t < 10^{-2}$ under isothermal conditions.

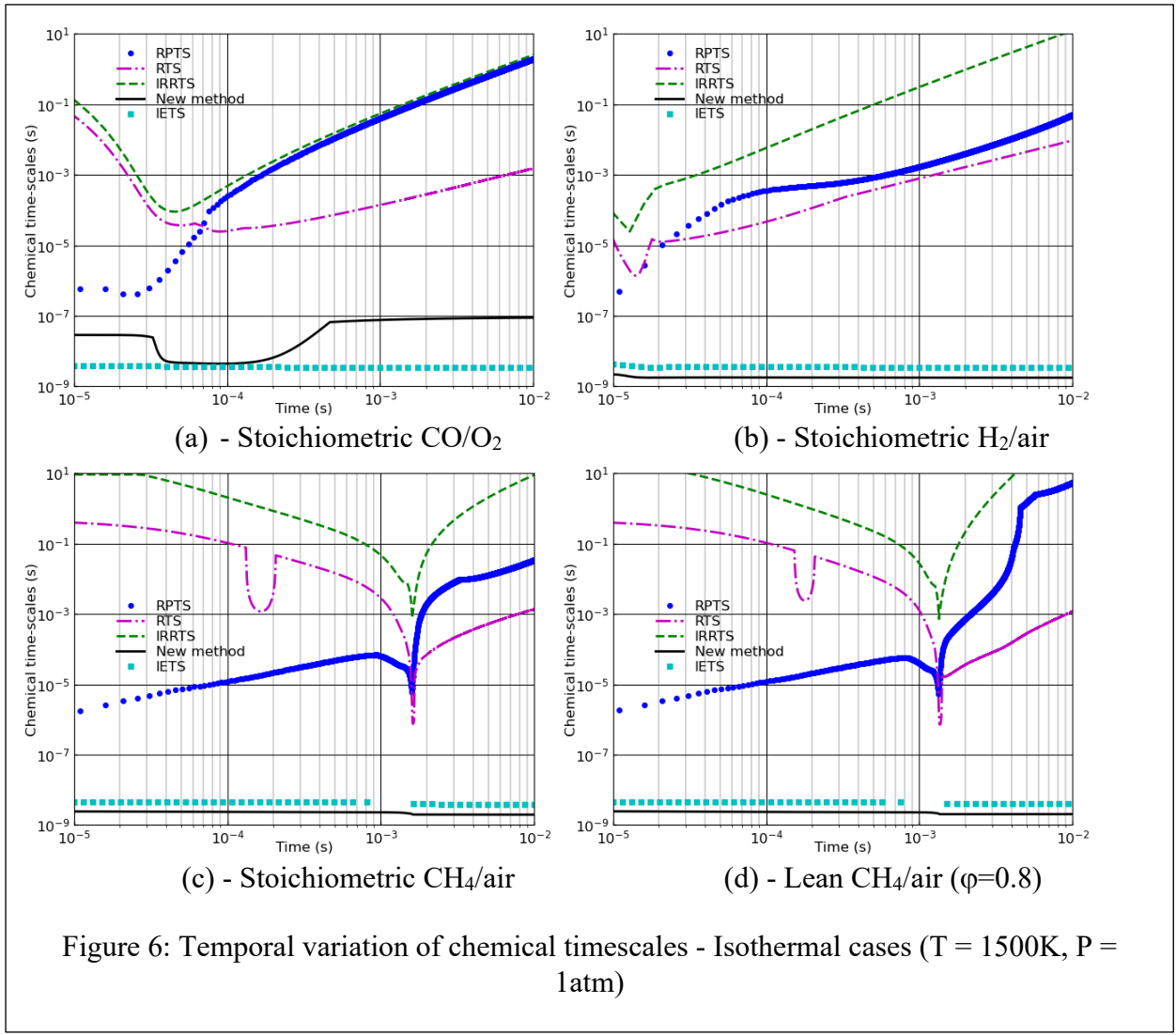

Figure 6: Temporal variation of chemical timescales - Isothermal cases (T = 1500K, P = 1atm)

For all the cases shown in Figure 6, there are some important common characteristics. The RTS, RPTS and IRRTS methods show that the timescales are at a minimum just near ignition of the fuel-oxidizer mixture. As the system approaches steady state the chemical time scale increase monotonically. As explained earlier, algebraic methods such as RTS and RPTS use the net production rate while the IRRTS method uses the net rate of progress to evaluate the timescales. As the system approaches steady state these terms approach zero and hence the timescale would tend to infinity. It is seen that for these methods, the temporal variation of the chemical timescales span almost three to four orders of magnitude as the system progresses from the initial condition to its final steady state. In contract, the timescales computed using the IETS method is practically constant throughout the time-evolution of the oxidation process and is about two to four orders of magnitude lower than the minimum timescales predicted by the algebraic methods (RPTS, RTS and IRRTS).

It is important to point out that the new method proposed in this work predicts timescales that share the characteristics of the IETS method. For the more interesting cases which show the complete combustion of $H_2$ and $CH_4$ (cases (b), (c) and (d)), the timescales predicted by the proposed method is less than a factor of two below those of the IETS method. As pointed out in Ref. [11], computing timescales using eigenvalue-based methods is computationally expensive and hence not practical for multidimensional turbulent reacting flow simulations. The proposed method computes the timescales while computing the net-production rate for each species with negligible additional computational cost. As mentioned above, timescales computed based on algebraic methods can have large temporal variation during the combustion process. The large temporal variation in the chemical timescales means large changes in the temporal variation of the local Damkohler number which can lead to large changes in turbulent chemistry models (like the EDC model) making the simulations numerically unstable. The chemical timescales (and hence the Damkohler number) predicted by the proposed method do no vary widely during the combustion process (as with the IETS methods), making multi-dimensional reacting flow CFD simulations more robust. Large values of chemical timescales predicted by the algebraic methods when the system approaches steady state can also lead to non-physical results. The authors in [11] report that high values of chemical timescales cause the Damkohler number to be very low which can lead to non-ignition of the flame while using the Eddy Dissipation Concept (EDC) model in turbulent reacting flow simulations. The discussion thus far shows that the proposed method of computing chemical timescales has the low computational cost of algebraic methods while having the same characteristics of nearly constant chemical timescales predicted by eigenvalue-based methods.

*4.3.2 Adiabatic cases:*

The time-evolution of a chemical system under constant pressure, adiabatic conditions is accomplished by the coupled solution of the system of equations describing the temporal variation of species evolution described by Eq (1) and the temperature using Eq (2). Figure 7 shows the temporal variation of the chemical timescales for various methods for the four cases discussed above, under adiabatic conditions. The temporal variation of the temperature of the system is also show (solid red line) to depict the transition of the mixture from its initial to the final state (temperature and species composition). It is seen that the timescales for the various methods under adiabatic conditions share the same qualitative characteristics as the timescales under isothermal conditions. The algebraic methods show a large temporal variation in the timescales (orders of magnitude), whereas the IETS and the new method proposed in this work show that the timescales vary less than a factor of two from the initial mixture state to the final state after undergoing complete combustion. The timescale predicted by the IETS method and the proposed method also differ by less than a factor of two at any given time during the combustion process for all the cases considered. Table 1 shows the chemical timescales for $CH_4$-air mixtures (stoichiometric and lean conditions) at steady-state for various methods. The timescales predicted by different algebraic methods differ widely as shown in Table 1 and the algebraic timescales are several orders of magnitude greater than the IETS method and the new proposed method for all the cases considered.

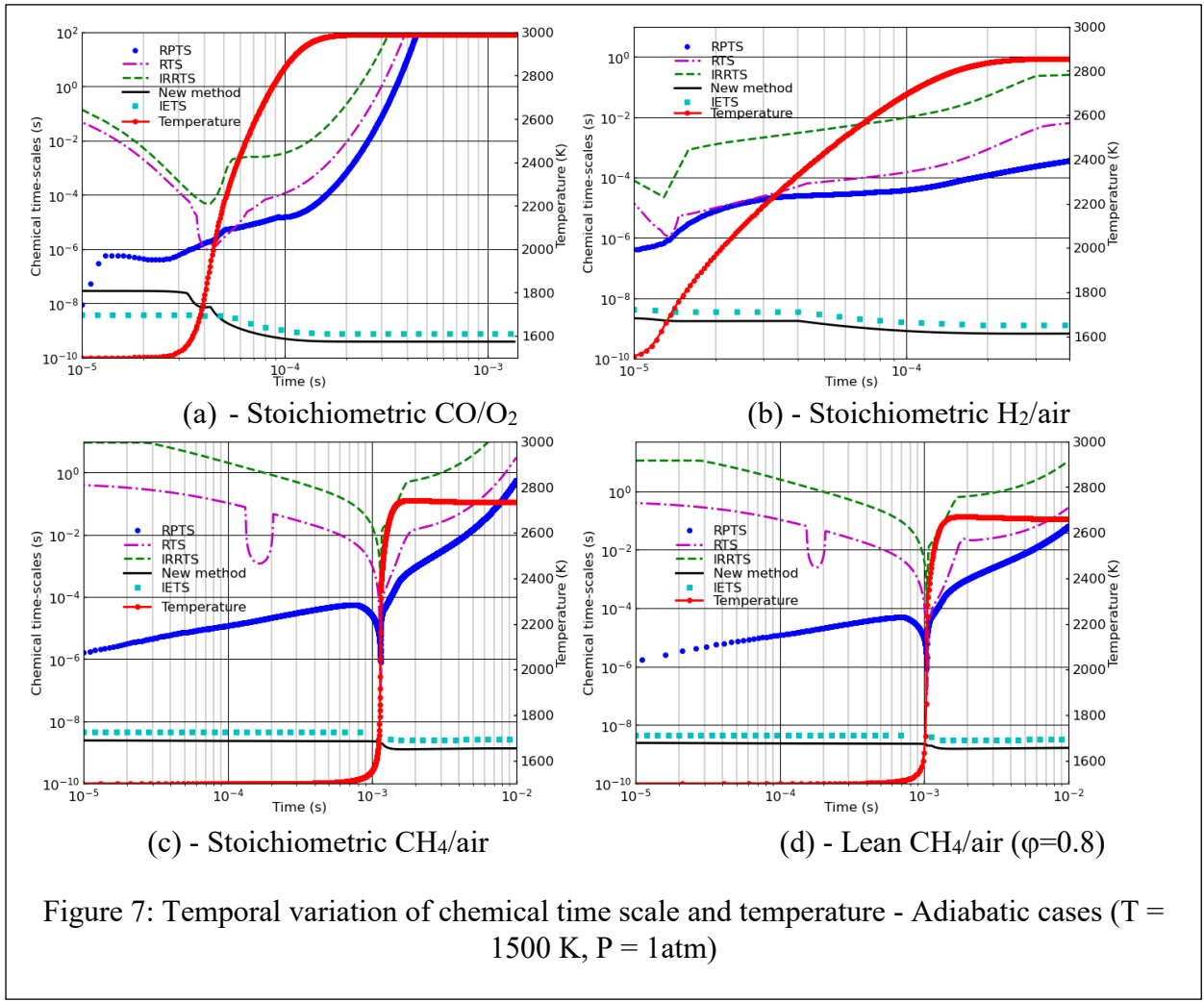

Figure 7: Temporal variation of chemical time scale and temperature - Adiabatic cases (T = 1500 K, P = 1atm)

Figure 7 (c and d) also shows that at t = $10^{-2}$ when steady-state temperatures have been reached, the chemical timescales predicted by the algebraic methods under lean conditions are about an order of magnitude lower than the stoichiometric $CH_4$/air mixtures. However, both the IETS and the proposed method show that under lean conditions the chemical timescale at steady state is higher than that under stoichiometric conditions, though the difference in timescales is only about 20%.

In addition to the numerical value of the chemical timescales it also useful to identify the species associated with each of the extremum values of $|\tau_k|$. For all cases of fuel-air combustion, namely, $H_2$-air and $CH_4$-air (stoichiometric and lean), the proposed method showed that $N_2$ had the longest chemical timescales throughout the combustion process (initial to final state). It is well-know that $N_2$ kinetics are orders of magnitude slower than C/H kinetics in combustion reactions. It was seen that the $N_2$ chemical timescales were on the order of tens of milliseconds as the combustion process proceeded for 1500 K < T < 1700 K. At steady state, when the temperatures were in excess of 2600K, the $N_2$ chemical timescales were a few milliseconds, as expected. The fastest chemical timescale during the initial stages of combustion was associated with the species 'NNH' whereas the fastest chemical timescale corresponded to $H_2O_2$ after the combustion process was complete and steady-state temperatures (> 2600 K) were reached.

*4.3.3 Near Ignition conditions in Natural gas engines:*

Hundreds of thousands of vehicles with natural gas engines are operating all over the world due their economic and environmental benefits. Natural gas engines generate almost no emissions of nitrogen oxides, particulate matter, volatile organic compounds, or carbon monoxide and have thus been widely used in a variety of medium and heavy-duty engine applications. Additionally, engines powered by natural gas cost significantly less than their gasoline and diesel counterparts and provide a pathway for a hydrogen economy. Given these benefits, there have been several modeling efforts for natural gas-powered engines [14-17]. We discuss the application of the proposed method in computing chemical timescales for the fuel-air mixture at conditions just prior to the spark (ignition). Typical values of the gas temperature and pressure prior to ignition is T = 750K and P = 25 atm.

Figure 8 shows the temporal variation of the chemical timescales and temperature under near ignition conditions in an internal combustion engine for both stoichiometric and lean $CH_4$-air mixtures.

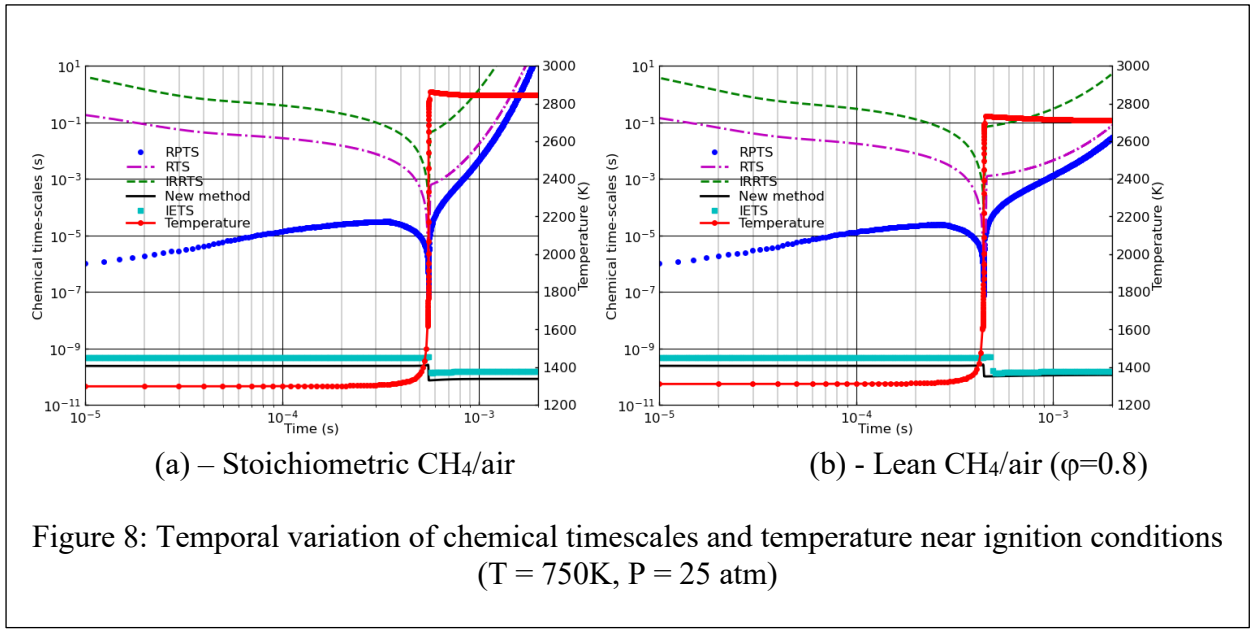

(a) – Stoichiometric CH$_4$/air  (b) - Lean CH$_4$/air ($\varphi$=0.8)

Figure 8: Temporal variation of chemical timescales and temperature near ignition conditions (T = 750K, P = 25 atm)

As with the earlier cases, the algebraic methods show a large variation (orders of magnitude) in the timescales during the combustion process and a monotonically increasing value of timescales after the temperature reaches a steady state. It is also seen that at t > $10^{-3}$ when steady-state temperatures have been reached, the chemical timescales predicted by the algebraic methods under lean conditions are about three orders of magnitude lower than the stoichiometric CH$_4$/air mixtures (see Table 1). This large variation in chemical timescales based on composition predicted by algebraic methods can lead to serious numerical instabilities in multi-dimensional reacting flow simulations where the equivalence ratios are expected to vary both spatially and temporally during the simulations. It is also seen that the proposed method shares the same characteristics as the IETS method with a minimal temporal variation in the timescales from initial to final (post combustion) conditions. The timescales of both the IETS and the proposed method under near-ignition conditions are about an order of magnitude lower than the chemical timescales at atmospheric pressure. This decrease in chemical timescale is expected on account of the fact that though the mixture is at a lower initial temperature (750K compared to 1500K), the pressure is twenty-five times higher. It is also noted that both the IETS method and the proposed method show that chemical timescales are a weak function of mixture composition (equivalence ratio) and that they differ by about 20% at steady state even at elevated pressures.

Furthermore, it was noted that $N_2$ had the longest chemical time constant throughout the combustion process (initial to final state) with the chemical timescales on the order of tens of milliseconds in the initial stages of the combustion process to a few milliseconds after steady state was reached (as with the case where the initial mixture was at 1 atm and 1500K). It was also noted that at higher pressures and lower initial temperatures (25 atm/750K), the fastest chemical timescales during the initial stages of combustion were due to $CH_2(s)$, as opposed to the species NNH for lower initial pressure and higher initial temperature (1 atm/1500K). At steady-state (post-combustion) temperatures, the fastest chemical timescales corresponded to $H_2O_2$ as with the lower pressure/higher initial temperature case.

| Case | $\tau$-IRRTS | $\tau$-RPTS | $\tau$-RTS | I $\tau$-ETS | $\tau$-Proposed |
|---|---|---|---|---|---|
| Adiabatic/Stoic | $>\approx 100$ | $>\approx 0.5$ | $>\approx 100$ | 2.6E-9 | 1.3E-9 |
| Adiabatic/lean | $>\approx 10$ | $>\approx 0.05$ | $>\approx 0.2$ | 3.2E-9 | 1.6E-9 |
| Engine/Stoic | $>\approx 10^4$ | $>\approx 30$ | $>\approx 100$ | 1.5E-10 | 8.3E-11 |
| Engine/lean | $>\approx 5$ | $>\approx 0.03$ | $>\approx 0.08$ | 1.47E-10 | 1.1E-10 |

Table 1: Chemical timescales (in sec) for $CH_4$-air mixtures (stoichiometric and lean conditions) at steady-state for various methods under adiabatic and engine pre-ignition conditions.

## 5 Conclusions

A new computationally efficient and numerically robust methodology to compute chemical timescales using detailed chemical kinetics was proposed in this work. The temporal variation of chemical timescales under a range of thermodynamic conditions (isothermal and adiabatic), fuel-air mixtures and initial conditions was studied using three different algebraic methods (IRRTS, RTS, RPTS) and the IETS eigenvalue method. The temporal variation of timescales predicted by these traditional methods was compared with the new algebraic method that addresses the deficiencies of the algebraic method and the eigenvalue-based methods. The effect of tight numerical tolerances on the predicted timescales was also studied. It was shown that very tight numerical tolerances are needed in algebraic methods, failing which, the predicted timescales are oscillatory in nature. The proposed method is computationally efficient as the algebraic methods but shows the same robust numerical and physical characteristics as the eigenvalue-based IETS method. All the algebraic methods studied in this work showed large temporal variation in the timescales with a monotonic increase in the chemical timescales even after the system had reached a steady state temperature. In contract, the eigenvalue based IETS method and the proposed method showed that the timescale of the system varied by less than a factor of two as the system evolved from the initial composition and temperature to the post-combustion temperature and composition. The IETS method and the proposed method also showed that the chemical timescale of the system was also constant after the system had reached steady-state temperatures. Quantitatively, the timescale predicted by the proposed methods was always within a factor of two of the IETS method. The proposed method also showed that the slowest time constant of the system was due to $N_2$ and that it was on the order of a few milliseconds at steady state for all the cases studied. The proposed method also showed that the fastest chemical time constant corresponded to NNH in the early ignition stages (1500 K< T < 1700K) at a pressure of 1 atmosphere and $CH_2$(s) at elevated pressures and initial lower temperatures. The fastest transient was associated with $H_2O_2$ at steady-state temperatures for both elevated pressures and atmospheric pressures. After steady state temperatures were reached, the algebraic method predicted timescales that were a strong function of mixture composition (equivalence ratio) and pressure. At atmospheric pressure, the lean mixtures showed timescales an order of magnitude lower than the stoichiometric mixture but were lower by as much as three orders of magnitude at elevated pressures. The IETS method and the proposed method both showed weak dependence on mixture composition and that the predicted timescales differed by about 20% both under atmospheric pressures and at elevated pressures seen in power generating equipment such as natural gas-powered engines. Use of the proposed method would enable the computation of chemical timescales for numerically robust multidimensional turbulent reacting flow simulations using detailed kinetics without the burdensome computational cost associated with eigenvalue-based methods and the numerical instability/non-physical results associated with algebraic methods.

**Acknowledgments** This material was based upon work supported by the U.S. Department of Energy, Office of Science, under Contract No. DE-AC02-06CH11357.